\documentclass[12pt,preprint]{aastex}

\usepackage{indentfirst}
\usepackage{floatflt,graphicx,rotating,subfigure}
\usepackage{longtable}
\usepackage{amsmath}
\usepackage{lscape}

\newcommand{\chandra}{{\em Chandra}}

\newcommand{\lya}{Ly$\alpha$}
\newcommand{\haroii}{Haro~11}
\newcommand{\hxi}{Haro~11~X-1}
\newcommand{\hxii}{Haro~11~X-2}

\newcommand{\chart}{ChaRT}
\newcommand{\kev}{keV}
\newcommand{\psf}{PSF}
\newcommand{\sherpa}{{\it Sherpa}}

\newcommand{\ciao}{{\it CIAO}}
\newcommand{\lyba}{LBA}

\newcommand{\lbg}{LBG}
\newcommand{\lae}{LAE}
\newcommand{\ulx}{ULX}
\newcommand{\bhb}{BHB}
\newcommand{\imbh}{IMBH}
\newcommand{\hmxb}{HMXB}
\newcommand{\hlx}{HLX}

\shorttitle{X-ray Binaries and Feedback in \lya\  Galaxies}
\shortauthors{Prestwich et al.}

\begin{document}

\title{Ultra-Luminous X-ray Sources in \haroii\ and the Role
  of X-ray Binaries in Feedback in \lya\ Emitting Galaxies}

\author{A.H. Prestwich}
\affil{Harvard-Smithsonian Center for Astrophysics, 60 Garden Street, Cambridge, MA 02138}
\author{F. Jackson}
\affil{Department of Physics and astronomy, University of Toledo, 2801 West Bancroft Street, Toledo, OH 43606}
\author{P. Kaaret \& M. Brorby}
\affil{Department of Physics and Astronomy, University of Iowa, Van Allen Hall, Iowa City, IA 52242}
\author{T.P. Roberts}
\affil{Department of Physics, University of Durham, South Road, Durham DH1 3LE, UK}
\author{S. H. Saar}
\affil{Harvard-Smithsonian Center for Astrophysics, 60 Garden Street, Cambridge, MA 02138}
\author{M. Yukita}
\affil{Johns Hopkins University, Homewood Campus, Baltimore, MD 21218, USA}


\begin{abstract}

Lyman Break Analogs (\lyba) are local proxies of high-redshift Lyman Break Galaxies (\lbg).    Spatially resolved studies of nearby starbursts have shown that  Lyman
continuum and line emission are absorbed by dust and that  the \lya\ is resonantly scattered by neutral hydrogen.   In order to observe Lya emission from star-forming regions,  some
source of feedback  is required to blow the neutral gas away from the
starburst to prevent scattering and allow the \lya\ emission to
escape.

 We show there are two X-ray point sources embedded in the diffuse emission of the Lyman Break Analog (LBA) galaxy \haroii.    CXOU J003652.4-333316 (abbreviated to \hxi) is an extremely luminous (L$_{X} \sim 10^{41}$ ergs s$^{-1}$),  spatially compact source with a hard X-ray spectrum.       We suggest the X-ray emission from \hxi\ is dominated by a single accretion source.  This might be an Active Galactic Nucleus (AGN) or a source similar to the extreme Black Hole Binary (BHB) M82 X-1.  The hard X-ray spectrum  indicates  \hxi\ may be a Black Hole Binary (BHB) in a low accretion state \citep{McClintock2006,Remillard2006}.  In this case, the very high X-ray luminosity suggests an intermediate mass black hole that could be the seed for formation of a supermassive black hole.

 Source CXOU J003652.7-33331619.5 (abbreviated \hxii) has an X-ray luminosity  L$_{X} \sim 5\times10^{40}$ ergs s$^{-1}$  and a soft X-ray spectrum (power law photon index $\Gamma \sim$2.2).   This strongly suggests that \hxii\ is an X-ray  binary in the ultra luminous state (i.e. an Ultra Luminous X-ray source, \ulx).   \hxii\ is  coincident with the star forming knot that is the source of the  \lya\ emission.   The association of a \ulx\ with \lya\ emission raises the possibility that strong winds from X-ray binaries play an important part in injecting mechanical  power into the Interstellar Medium (ISM), thus blowing away neutral material from the starburst region and allowing the \lya\  to escape.    We suggest that  feedback from X-ray binaries may play a significant role in allowing \lya\ emission to escape from galaxies in the early universe.

\end{abstract}

\section{Introduction}

The youngest  galaxies in the early universe are strong sources of 
\lya\ and Lyman continuum emission from young, massive stars (\cite{Steidel1999},  and references therein, \cite{Shapley2006,Nilsson2007,Mallery2012}).    These are the systems thought to be responsible for reionizing the early universe \citep{Haiman1999, Dijkstra2007, Mesinger2008, Fontanot2012}.      Two techniques are used to discover large numbers of galaxies at high redshifts: the direct detection of \lya\ emission in a narrow band filter \citep{Hu1996,Gronwall2007,Ouchi2008,Mallery2012} and the Lyman Break technique \citep{Steidel1999,Steidel1995, Steidel1996}.    The Lyman Break technique makes use of the fact that star forming galaxies have spectral energy distributions which exhibit a large drop in flux at wavelengths shorter than the Lyman limit in the rest frame of the galaxy \citep{Steidel1995}.   The Lyman break is likely due to a combination of the intrinsic spectral shape of the young stellar population and absorption by neutral gas.   Deep imaging of objects above and below the Lyman limit  can identify objects that disappear at shorter wavelengths: this ``dropout" is likely due to the Lyman break  \citep{Steidel1995}.   Objects discovered via their \lya\ emission are referred to as \lya\ Emitters (\lae s), and those identified  via the Lyman break as  Lyman Break Galaxies (\lbg).

Despite the importance of Lyman emission,  both for high redshift surveys and for understanding galaxy evolution, the details of how
Lyman emission escapes from a galaxy are not well understood.   This is because  \lya\ line and continuum emission cannot be spatially resolved in high redshift systems.  To address this problem,  \cite{Heckman2005} and \cite{Hoopes2007}  derived a sample of compact, UV bright, low-redshift galaxies using GALEX and the Sloan Digital Sky Survey (SDSS).  These galaxies have properties that are very similar to \lbg, including size, UV luminosity, surface brightness, mass, star formation rate and metallicity  \citep{Hoopes2007,Heckman2005}.    They are known as Lyman Break Analogs (\lyba) and are used as proxies of high-redshift \lbg.  Spatially resolved studies of local starburst galaxies have shown that  Lyman
continuum and line emission is absorbed by dust, reducing the
\lya\ line strength from values predicted by simple models of HII
regions \citep{Kunth2003,Mas-Hesse2003,Ostlin2009,Scarlata2009}.   In addition, the \lya\ is resonantly scattered by neutral
hydrogen surrounding the starburst region (\cite{Neufeld1991,Hayes2010},\cite{Verhamme2012} and references therein)   It appears that some
source of feedback  is required to blow the neutral gas away from the
starburst to prevent scattering and allow the \lya\ emission to
escape \citep{Wofford2013, Orsi2012}.  The most obvious source of mechanical power is supernovae and stellar winds from the starburst region creating ``super-bubbles'' \citep{Tenorio-Tagle1999,Hayes2007,Heckman2011}.

\begin{figure}[ht!]
     \begin{center}

 \includegraphics[scale=1.0]{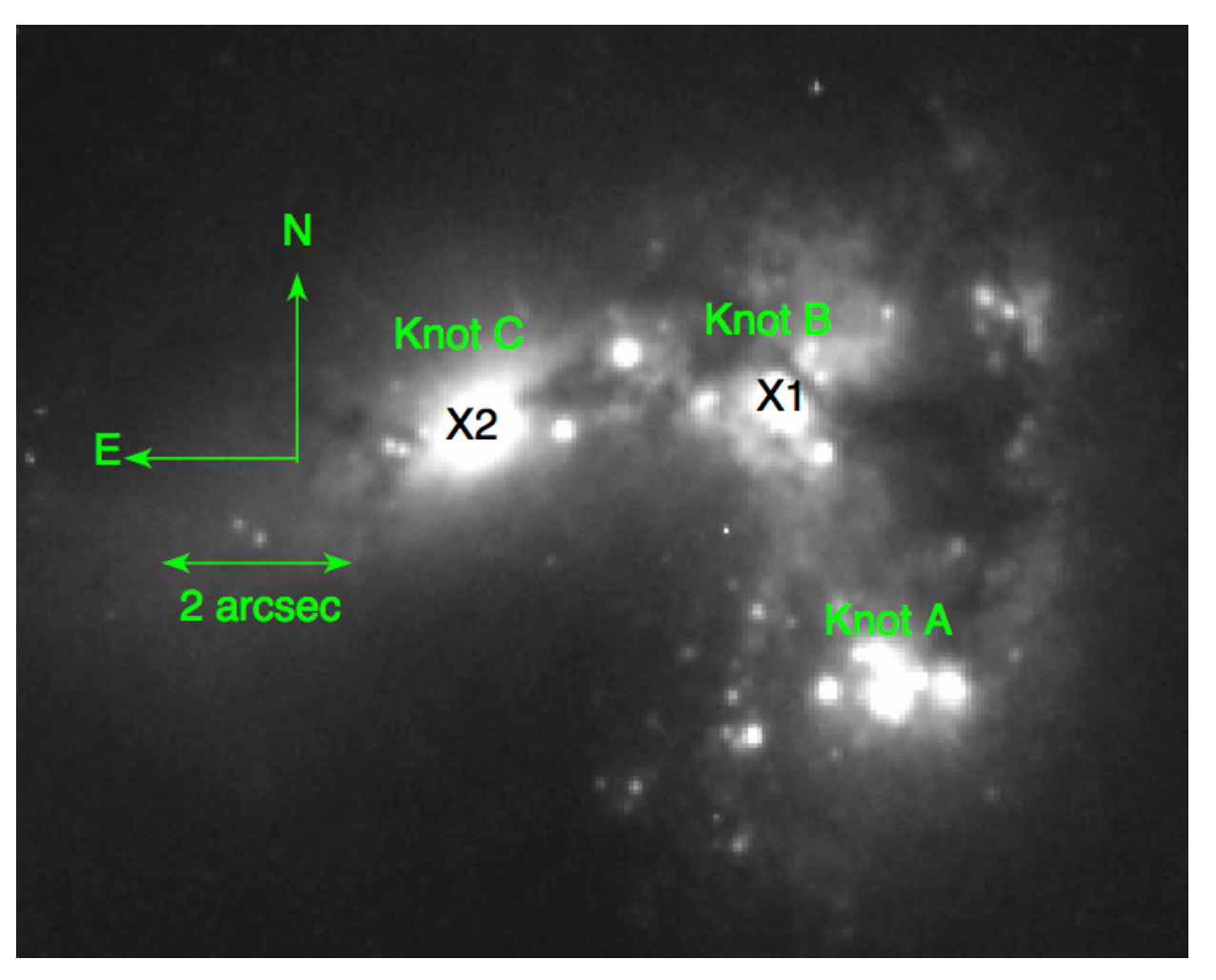}
 
  \end{center}
  \caption{Image of \haroii\ showing the three star forming knots identified by \cite{Vader1993}.   Also shown are the locations of the two bright X-ray point sources \hxi\ and \hxii.   The image was taken by the Hubble Space Telescope (HST) Advanced Camera for Surveys Wide Field Camera in F435W ban and downloaded from the Hubble Legacy Archive. }%
   \label{fig:optical}
\end{figure}

 \haroii\  is a famous compact dwarf galaxy undergoing an intense starburst that has been intensively studied at all wavelengths \citep[see][and references therein]{Adamo2010}.    The starburst region is resolved into three distinct knots, labeled A, B and C by \cite{Vader1993} (see Figure~\ref{fig:optical}).   It is a  \lya\ emitter \citep{Kunth2003} and is in the \cite{Hoopes2007} sample of  \lyba.  In addition,   \haroii\  is the only known example of a local Lyman continuum emitter \citep{leitet2011}.   The \lya\ emission is centered on a massive star forming region (Knot C).   \cite{Grimes2007} studied the global X-ray emission of \haroii.   They  showed that the  X-ray emission is spatially extended and that the integrated spectrum of the entire galaxy  is composed
of a diffuse thermal  component associated with hot gas and a power
law component likely associated with the X-ray binary population of
the galaxy \citep{Grimes2007}.     

 In this paper, we study the X-ray properties of two point sources in \haroii.      CXOU J003652.4-333316 (abbreviated to \hxi) is an extremely luminous (L$_{X} \sim \times10^{41}$ ergs s$^{-1}$) source with a hard X-ray spectrum which is coincident with Knot B.   We suggest the X-ray emission from \hxi\ is dominated by a single accretion source.  This might be an Active Galactic Nucleus (AGN) or a source similar to the extreme Black Hole Binary (BHB) M82 X-1.  The hard X-ray spectrum  indicates  \hxi\ may be a Black Hole Binary (BHB) in a low accretion state \citep{McClintock2006,Remillard2006}.  If this is the case,  the very high X-ray luminosity suggests an Intermediate Mass Black Hole (IMBH).      If this interpretation is correct, \hxi\ is an extremely unusual object.   Finally, we note that an IMBH in a young galaxy may form the seed for a supermassive black hole - we may be witnessing the birth of an AGN in \haroii.    
 
 Source CXOU J003652.7-33331619.5 (\hxii) is coincident with Knot C, also at the center of the \lya\ emission .  The X-ray luminosity of \hxii\  and the soft X-ray spectrum  suggests it is a \ulx.  \ulx\ are known to have powerful jets and winds that impact their environment \citep{Abolmasov2011,Abolmasov2007,Pakull2002}.  The fact that a \ulx\ is coincident with Knot C raises the possibility that winds from the compact object may play a significant role in sweeping up the neutral gas and allowing the \lya\ to escape.    We suggest that  feedback from X-ray Binaries may be significant in the early universe \citep{Pacucci2014,Fragos2013,Justham2012}. 
 
We adopt a distance D=84.0 Mpc to \haroii\ resulting in a scale of 407 parsec per arcsecond (NASA Extrgalactic Database).

 \section{X-ray Observations}
\label{sc:xray}

\haroii\  was observed with \chandra\ ACIS-S array for 54 ks on  28 Oct 2006, obsid 8175.   Data were reprocessed using CIAO version 4.5 and CALDB 4.5.5.1.

\begin{figure}[ht!]
     \begin{center}

\includegraphics[scale=0.7]{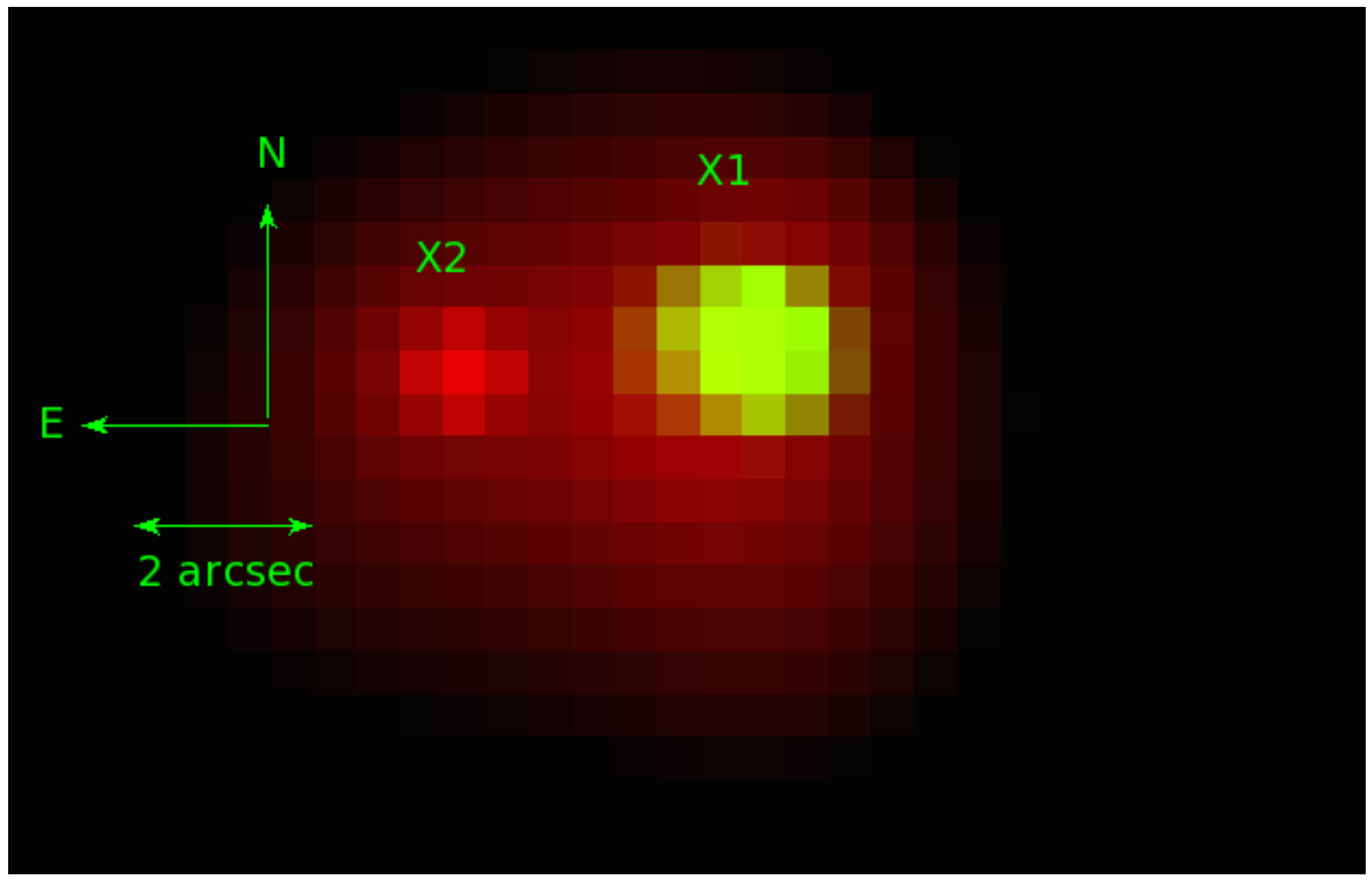}
  
    \end{center}
    \caption{Two band (0.3-1 keV=red and 3-5 keV=green) false color image of the X-ray emission.  The two point sources, \hxi\ and \hxii\ are shown.} 
   \label{fig:Xray_2col}
\end{figure}

\begin{figure}[ht!]
     \begin{center}

 \includegraphics[scale=1.0]{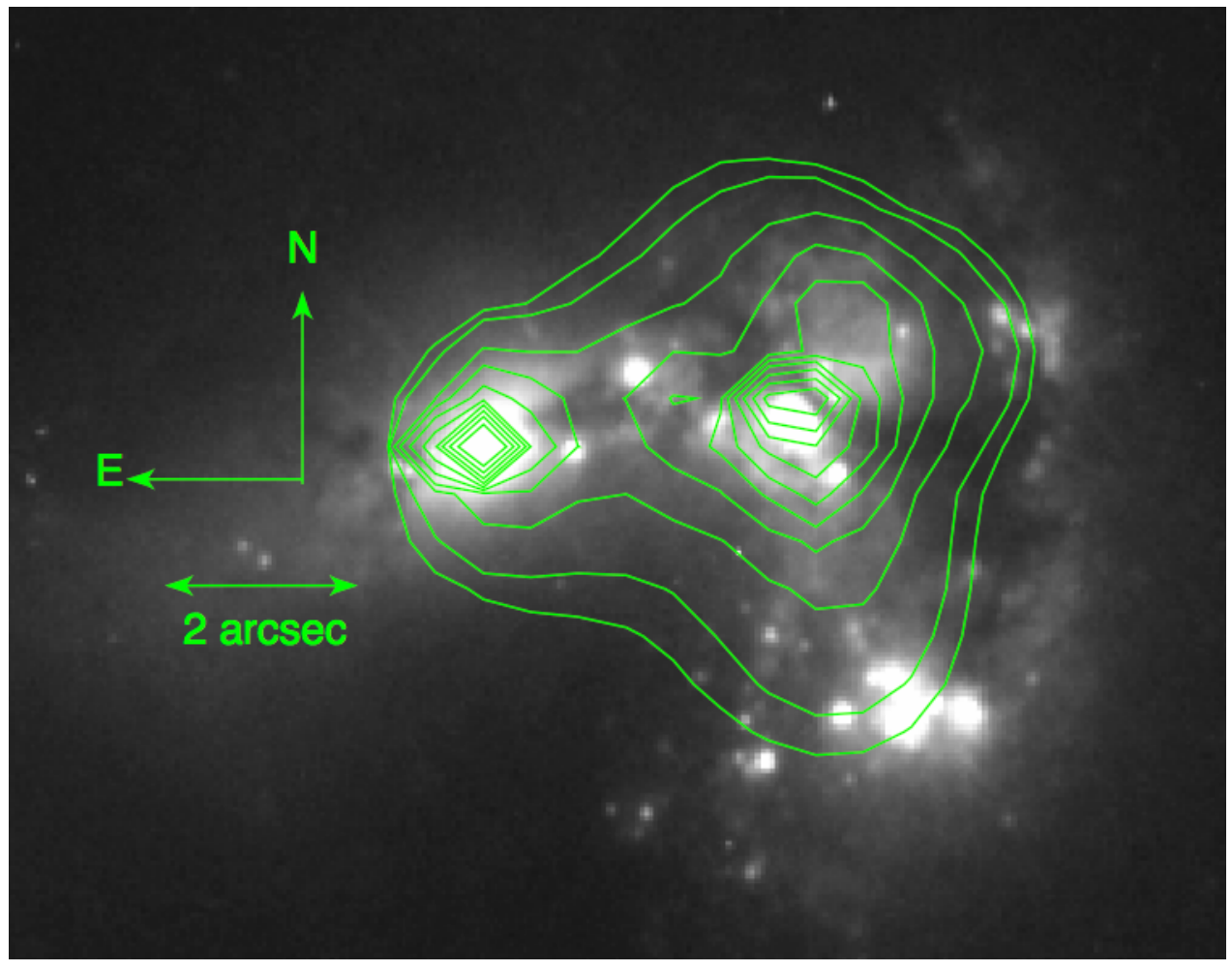}
  
    \end{center}
    \caption{ X-ray contours (0.3-5 keV)  superimposed on the HST optical image.}
   \label{fig:hst_Xray}
\end{figure}

\begin{figure}[ht!]
     \begin{center}
 \includegraphics[scale=1.0]{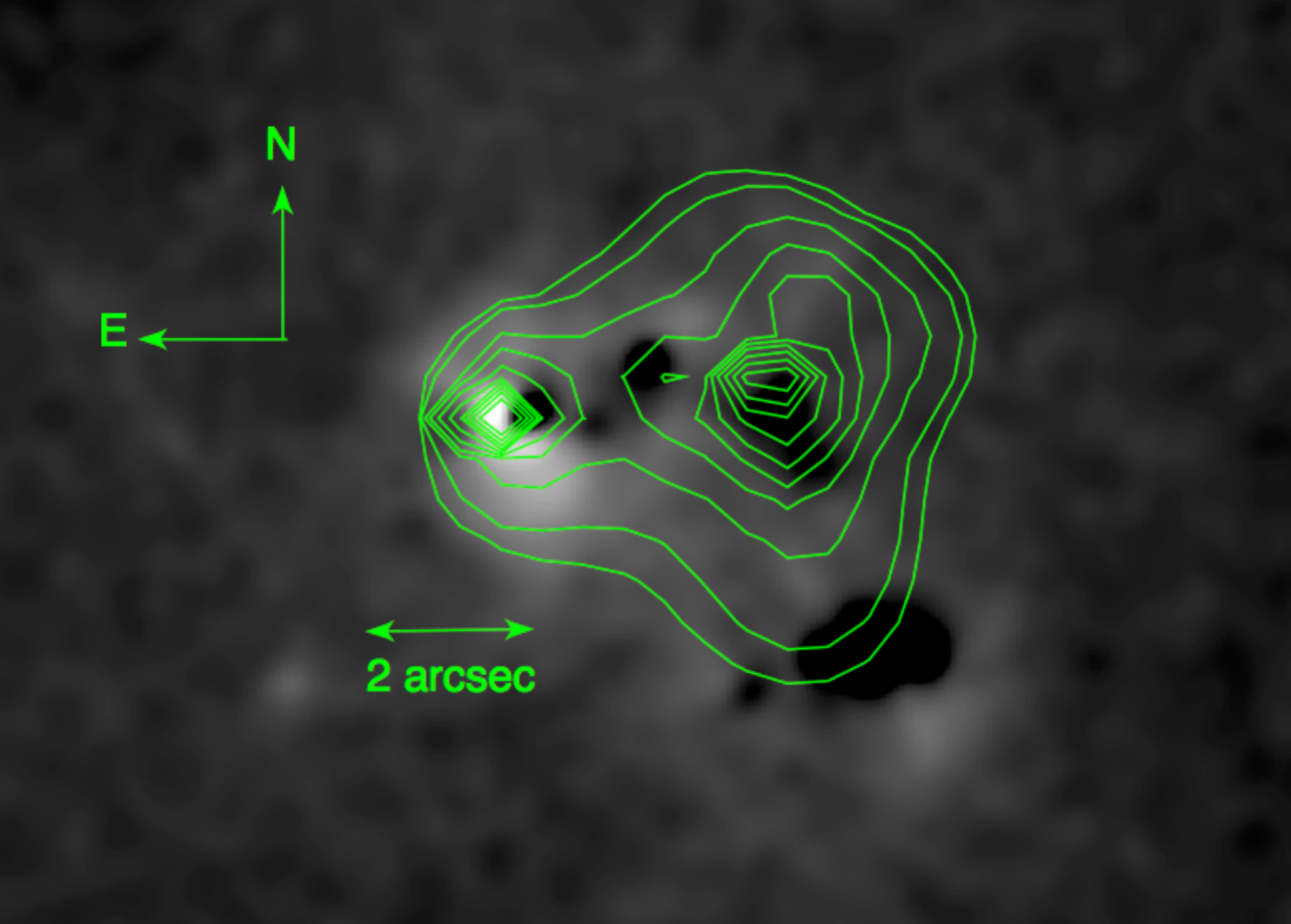}
    \end{center}
    \caption{ X-ray contours (0.3-5 keV) superimposed on the \lya\ line emission \citep{Ostlin2009}.   The \lya\ image is continuum subtracted.  The ULX \hxii\ is associated with Knot  C and the  \lya\  line emission.   \hxi\ is associated with Knot B.}
   \label{fig:Lya_Xray}
\end{figure}

A two-band  (0.3-1 keV=red and 3-5 keV=green)  false color X-ray image of \haroii\  is shown in  Figure~\ref{fig:Xray_2col}.   Two bright point sources are visible, one of which (\hxi)  has a hard spectrum and is clearly visible in the 3-5 keV band.  There is a second point source visible (\hxii) to the east of \hxi.  There is also extended soft X-ray emission.  Figure~\ref{fig:hst_Xray} shows the location of the X-ray emission (red contours) relative to star forming Knots A, B, and C discussed by \cite{Vader1993, Kunth2003, Adamo2010}.      The source  \hxi\ is associated with Knot B and \hxii\ is associated with Knot  C. Figure~\ref{fig:Lya_Xray} shows the \lya\ emission from \cite{Ostlin2009} with X-ray contours superimposed.  The bulk of the \lya\ line emission is associated with Knot C and \hxii.

\subsection{Spectral Analysis of the Point Sources}
\label{sec:spec}

Spectra were extracted for both the point sources (\hxi\ and \hxii)  in the range 0.3-10 keV  using the \ciao\ tool {\it dmextract}.    Spectra were also extracted from annular background regions (inner radius=$r_1$, outer radius=$r_2$) centered on the sources.   The source and background extraction radii were chosen by visual inspection of the sources.    Details of the extraction regions used are in Table~\ref{tab:spec_fit}.  \sherpa\ was used to fit the spectra in the 0.5-8 keV range.  The spectra were grouped with a minimum of 15 counts per bin, and  fit with an absorbed power law (\sherpa\ syntax xswabs*powlaw1d).   For both  fits, the absorption was allowed to vary, but not to drop below the galactic value of 1.88$\times$10$^{-20}$ cm$^{-2}$ \citep{Dickey1990}.    Both sources are well fit with an absorbed power law.  The best fit spectrum of \hxi\  is hard, with a power law slope   $\Gamma$=1.2$\pm 0.2$ (Table~\ref{tab:spec_fit}.).


\begin{table}[!h]
\centering
\caption{Extraction Regions and Best fit spectral parameters.}
\begin{tabular}{lcc}
\hline
\hline
&\hxi &\hxii \\
\hline
RA & 00:36:52.42 & 00:36:52.70\\
Dec & -33:33:16.95  & -33:33:16.95\\
Extraction radius (arcsec) & 1.1 & 1.1\\
Background annulus, $r_1$-$r_2$(arcsec)& 3.6-4.8 & 1.2-2.3\\
Source Counts &472 & 232\\
Background Counts & 291 &60\\
Model & PL & PL \\
N$_H$ ($\times10^{20}$ cm$^{-2}$)  &  18.8$^{+14.6}_{-12.7}$ &14.8$^{+9.9}_{9.9}$  \\
$\Gamma$  &1.2$^{+0.2}_{-0.2}$  &2.2$^{0.4}_{-0.4}$ \\
PL Flux$^{1}$&11.8 $^{+3.7}_{-4.5}$& 5.5$^{+3.7}_{-2.3}$  \\
\hline\\
X-ray Luminosity$^{2}$& 9.9$^{+8.4}_{-1.3}$& 4.7$^{+6.1}_{-1.0}$ \\
\hline\\
$\chi^2$/DOF &19.4/24 & 5.61/13\\
\hline\\
\multicolumn{3}{l}{$^{1}$ F$_{X}$(0.3-8.0 keV)$\times$10$^{-14}$ ergs s$^{-1}$ cm$^{-2}$}\\
\multicolumn{3}{l}{$^{2}$ L$_{X}$(0.3-8.0 keV)$\times$10$^{40}$ ergs s$^{-1}$}, (not corrected for absorption)\\
\hline\\
\end{tabular} 
\label{tab:spec_fit}
\end{table}


\begin{figure}[ht!]
     \begin{center}

\includegraphics[width=0.6\textwidth]{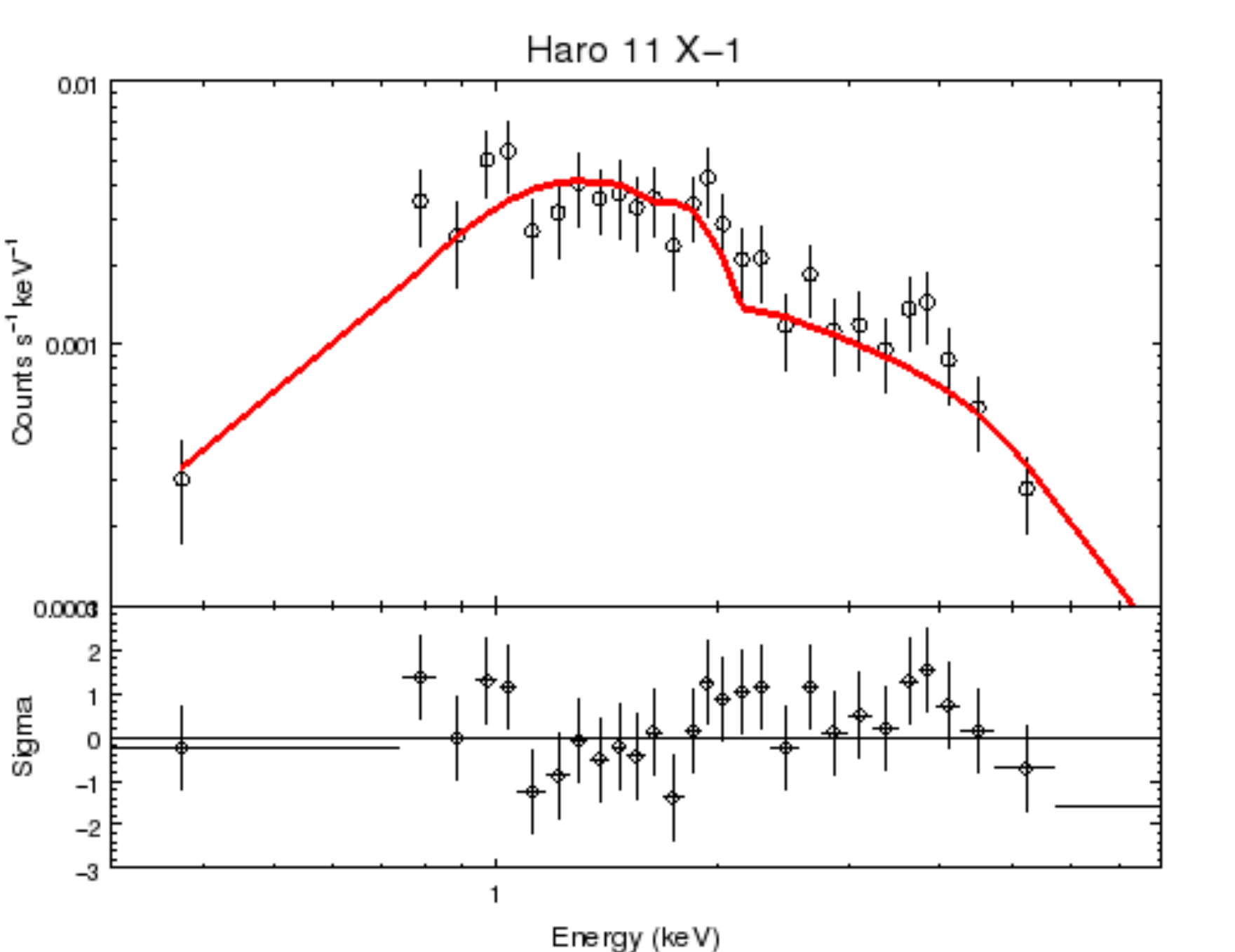}
 \includegraphics[width=0.6\textwidth]{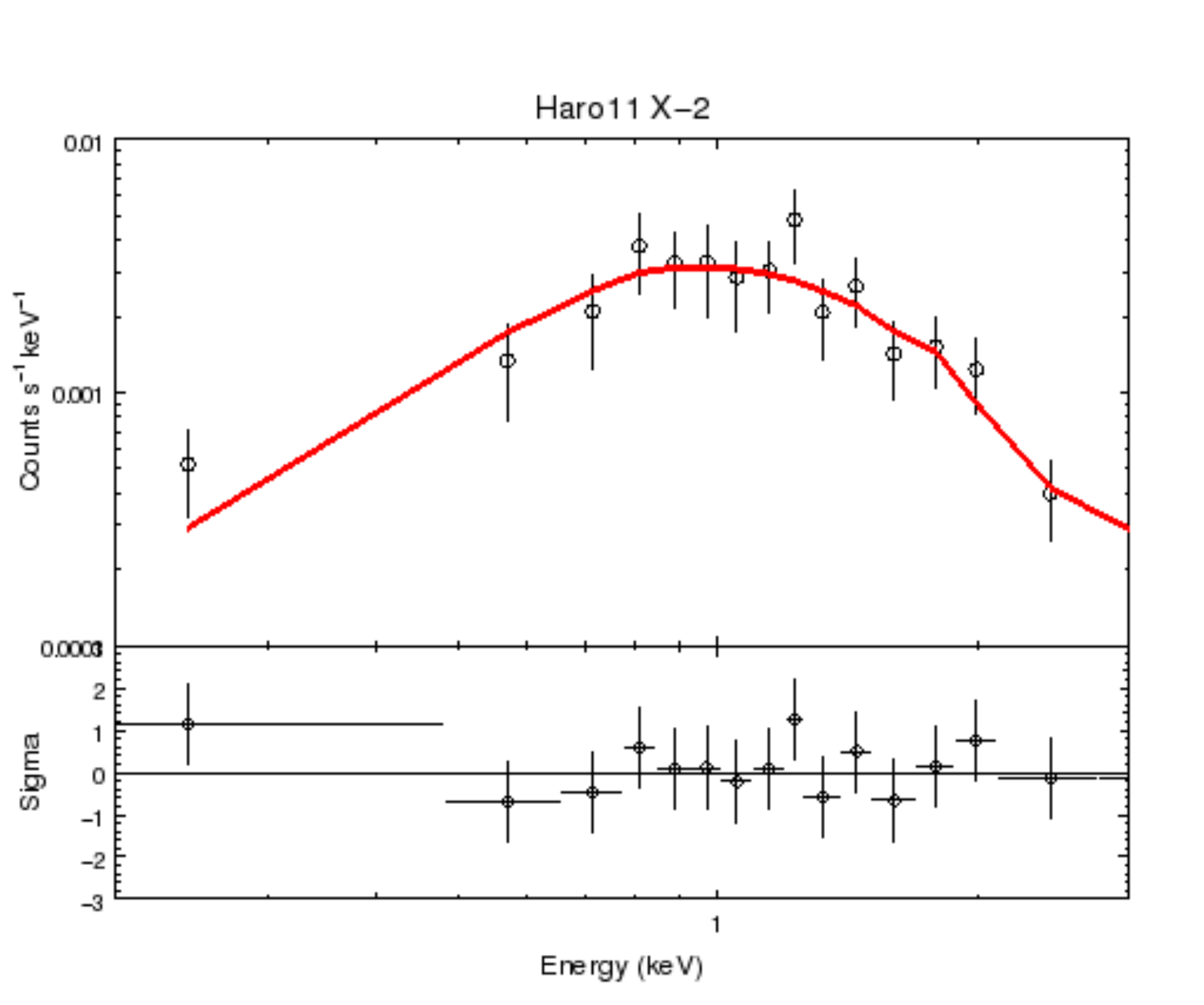}
   
    \end{center}
    \caption{Best fit X-ray spectra and residuals for the two bright point sources.  The top panel shows \hxi\ and the bottom panel \hxii.   The best fit models are shown in red.  Best fit models are shown in Table~\ref{tab:spec_fit}.}
   \label{fig:Xray}
\end{figure}


\section{The Nature of \hxi}

\hxi\ is an unusual source.   Most of the emission appears to come from a spectrally hard unresolved source with X-ray luminosity L$_X\sim10^{41}$ ergs s$^{-1}$.   This  is very high for a single X-ray binary,  and is comparable to the intermediate mass black hole candidate HLX-1 \citep{Servillat2011}.    This source is coincident with star forming Knot B.   

\subsection{The Spatial Extent of \hxi}

 The X-ray luminosity of \hxi\ is very high for a single X-ray binary and has a hard spectrum.  We therefore consider whether it is powered by an AGN, one or more X-ray binaries  or other high energy processes  associated with star formation (e.g. such as outflowing gas from a starburst region \citep{Yukita2012}, inverse Compton scattering of infrared photons off of relativistic electrons \citep{Hargrave1974,Rieke1980}, or synchrotron from very high energy electrons \citep{Lacki2013}).    In this section, we put observational constraints on the X-ray spatial extent of \hxi.  If the source is extended, we can definitively say that not all of the X-ray flux comes from a single accretion source.  We note that at the distance of \haroii, one arcsecond corresponds to approximately 400 pc.  Failure to detect spatial extent does not rule out a contribution from an extended component. 

We compare the radial profile of \hxi\  in the 3-5 \kev\ band with a monochromatic  Point Response Function (\psf) at 4 \kev (generated using \chart).   The 3-5 \kev\ band was chosen because the thermal hot gas (kT$\sim$1 \kev) contributes very little flux above 3 \kev\ and the flux from \hxi\ drops off rapidly above 5 \kev.   The source and \psf\ images were binned at 0.25 arcseconds per pixel.  We used \sherpa\  2-dimensional fitting routines  to obtain the best fit \psf.  The \psf\ generated from \chart\ was used as the convolution kernel and a 2-D Delta function (i.e. a point source) used as the model.   The best fit scaled \psf\ was obtained.    We then obtained the radial profile of the \psf\ and fit it with a 1-D Gaussian to derive a 1-D \psf.   Parameters of the 1-D \psf\ are shown in Table~\ref{tab:psf}.  The radial profile of \haroii\ was fit with a model comprising the 1-D \psf\ plus a constant background.  The background was calculated to be 0.2 counts per arcsecond$^{2}$.  This is a combination of instrument background (see the {\it Proposers Observatory Guide V. 16}) plus a contribution from the extended emission from \haroii.   The extended component is composed of thermal emission plus a power law component.  It is the power law that contributes a small amount of flux at 4 keV.  

\begin{figure}
\includegraphics[scale=0.8]{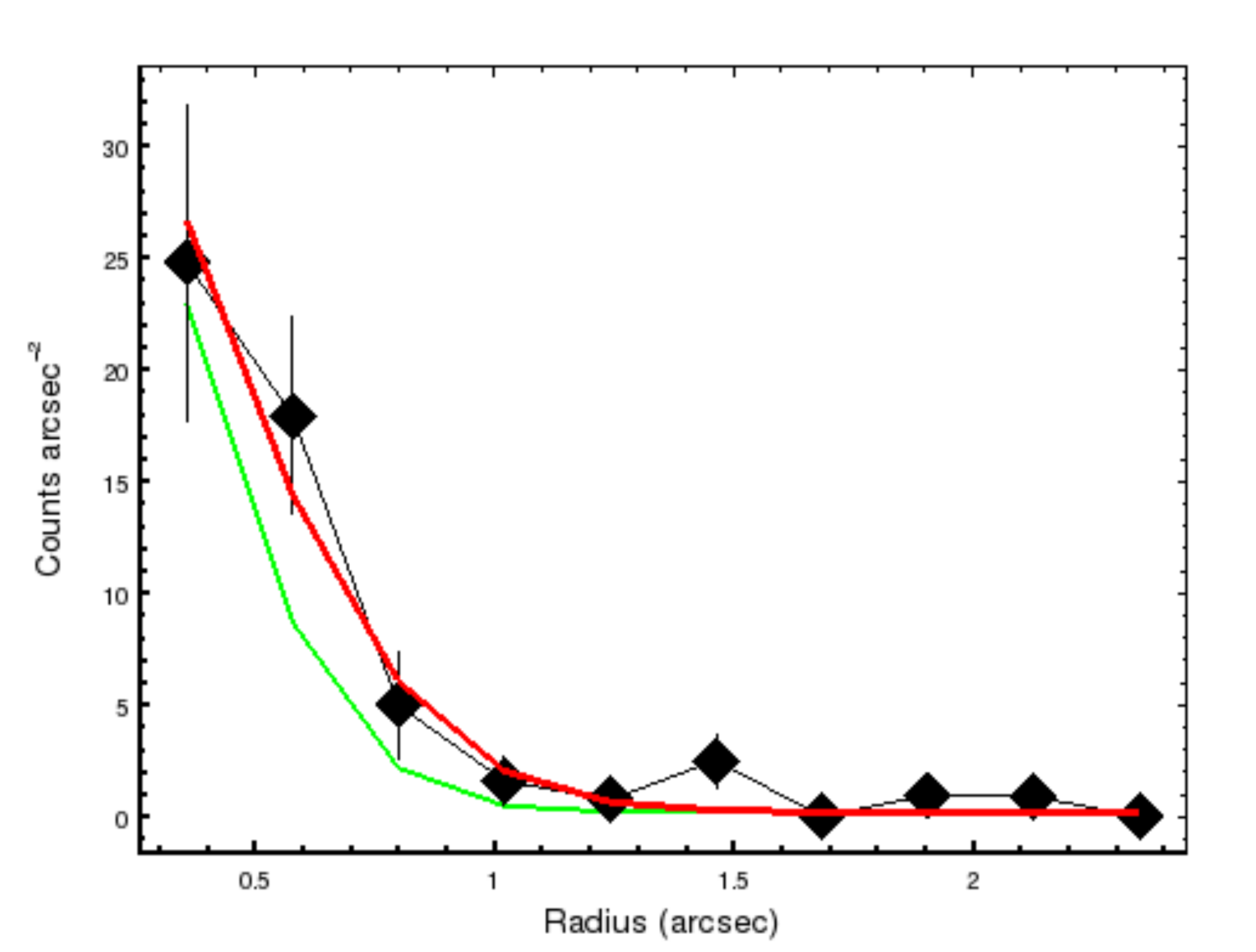}
\caption{Radial profile of \hxi\  in the 3-5 keV band (black) compared to the best fit model.  The model was generated by allowing the FWHM of the PSF and the background to vary.   A 4keV point response function generated using \chart\ and scaled to the data is also shown in green. The radial profile of \hxi\ is slightly broader than the PSF.  Parameters of the best fit spatial model are shown in Table~\ref{tab:psf}}
\label{fig:rad_prof_4kev}

\end{figure}

We initially fixed all the model components (background, Gaussian width, amplitude and position) with their default values and used \sherpa\ to calculate the  $\chi^{2}$ and $\chi^{2}$ per degree of freedom (the reduced $\chi^{2}$).   Despite the fact that no parameters were allowed to vary, the \psf\ plus background was an acceptable fit to the data (reduced  $\chi^{2}$=0.28).   In order to test the hypothesis that the source may be slightly extended, we allowed the FWHM and background to change during a fit.  This  improved the fit slightly (Table~\ref{tab:psf}) changing  $\chi^{2}$ from 1.93 to 1.46 (final reduced $\chi^{2}$=0.18.  The best fit Gaussian FWHM is 1.0  arcseconds, compared to 0.76 arcseconds for the \psf.   The radial profile of the compact source, the best fit and the original keV  \psf\ are shown in Figure~\ref{fig:rad_prof_4kev}.  We conclude that  \hxi\ is slightly extended beyond the \psf.     The extended component contributes $\sim$4 counts to the 3-5 keV flux (4\% of the total  3-5 keV flux)  

\begin{table}[!h]
\centering
\caption{Best fit spatial components.}
\begin{tabular}{ccccc}
\hline
\multicolumn{2}{c}{Gaussian} &  &\\\cline{1-2}
FWHM &  Amplitude & background & $\chi^{2}$/dof  & Note\\
arcsec &  & cts arcsec$^{-2}$ & & \\
\hline\\
0.76 & 41.75 &   0.2 & 2.5/9 & parameters from PSF, no fit \\
1.0 & 41.75 & 0.51 & 1.4/7 & background and FWHM allowed to change\\
\hline
\end{tabular} 
\label{tab:psf}
\end{table}

\subsection{Physical Interprtation of \hxi} 

We suggest the X-ray characteristics of \hxi\  (hard spectrum, compact source, high luminosity)  are best explained if the X-ray emission is dominated by a single accretion source.   We note that the radial profile fit is improved slightly with a background component in addition to the instrumental background.  This background is likely due to  high energy processes in starburst regions  (e.g. outflowing gas from a starburst region, inverse Compton scattering of infrared photons off of relativistic electrons and synchrotron from very high energy electrons) and unresolved X-ray binaries.   

\hxi\ also appears to be coincident with a  unresolved/compact  HI absorption feature recently reported by \cite{McHattie2014}.     The 21cm continuum emission  peaks at the same location \citep{McHattie2014}.  

\subsubsection{Low Luminosity AGN}
\label{sec:agn}
To our knowledge, there have been no reports of a LLAGN in \haroii.  The optical line ratios are generally consistent with HII regions \citep{Vader1993,James2013}.  The mid infrared emission lines of low metallicity starbursts are discussed by \cite{Hao2009}.  They find that the [OIV]]25.9$\micron$/[SIII]33$\micron$ and  [Ne III]15.56$\micron$/[Ne II]12.81$\micron$ ratio is a good discriminator between AGN and starbursts.  \haroii\ is well within the starburst region using these diagnostics.   Finally we note that \cite{Heisler1998} observed \haroii\ at 2.3 GHz with the Parkes-Tidbinbilla interferometer, which is sensitive to high surface brightness cores (spatial extent $<$ 0.1 arcsecond and T$_{b}> 10^5$K).  No source was detected. 

We cannot rule out a LLAGN whose optical and IR signature is diluted by the intense star formation occurring in Knot B \citep{Jackson2012}.  We can put an upper  limit on the black hole mass using the (5$\sigma$) upper limit to the radio  flux from \cite{Heisler1998}  and  the X-ray-radio black hole ``fundamental plane'' \citep{Merloni2005}.      Assuming a 2.3 GHz  GHz flux of $<$ 3.8 mJy, and further assuming that any compact core has a flat spectrum so that $S_{5GHz}\sim S_{2.3GHz}$,   we derive a black hole mass of $<$5$\times 10^{7}$ M$_{\odot}$.    

\cite{Adamo2010} estimate that the star formation in Knot B has been on-going for  $\sim$3.5 Myr.  If there is a supermassive black hole buried in Knot B, it is unlikely to have formed in the current star formation event, which is probably triggered by a galaxy merger.  The seed black hole must have been present in one of the galaxies prior to the merger, and it is now being fueled by the starburst and merger process.  

\subsubsection{One or more X-ray Binaries: an Intermediate Mass Black Hole?}
\label{sec:xrb}

\begin{figure}
\includegraphics[scale=1.5]{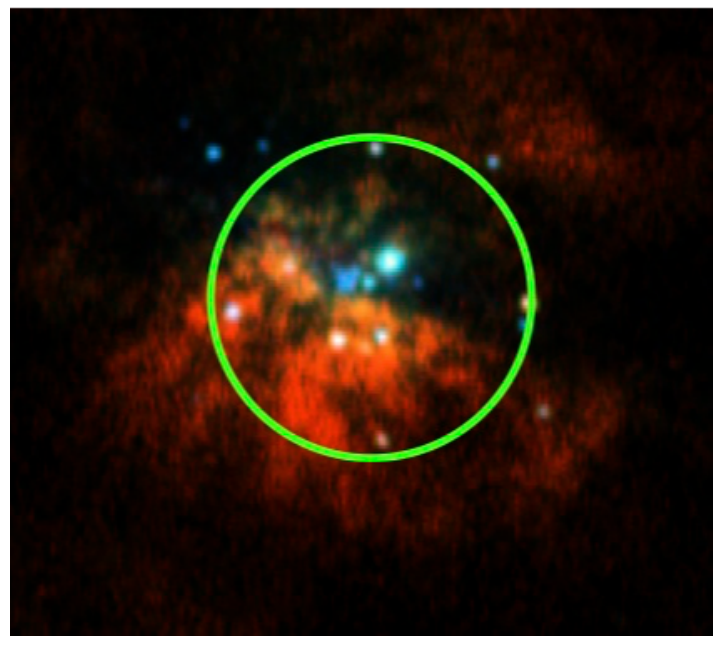}
\caption{ \chandra\ image of the starburst core of M82.   The brightest source is M82 X-1.  The compact source \hxi\ may comprise a very luminous source similar to M82 X-1 plus residual emission from the surrounding starburst.  The green circle shows the area encompassed by a 1 arc second aperture of the distance of \haroii}
\label{fig:M82}
\end{figure}

One or more X-ray binaries can naturally explain both the energetics and hard X-ray spectrum of \hxi.   In this context, it is useful to compare the X-ray source population of \haroii\ to that of  the nearby startburst galaxy M82.   The central starburst region in M82 contains $\sim$22 sources within a 2.5$\times$2.5 kpc area \citep{Griffiths2000}.    The integrated flux from the point sources is dominated by M82 X-1,  a single bright variable source \citep{Kaaret2001,Matsumoto2001}.    In the low state, M82 X-1 contributes 30\% of the flux from point sources in M82.   In the high state M82 X-1 dominates the point source flux.   It seems plausible that the compact source \hxi\ is similar to the starburst region of M82, except that \haroii\ is significantly further than M82 (84 Mpc vs. 3.7 Mpc) and hence individual sources are not detected and resolved.   This is illustrated in Figure~\ref{fig:M82}, which shows the area encompassed by a 1 arcsecond aperture at the distance of \haroii.  

The very hard spectrum is consistent with \hxi\ being dominated by a single source.   \cite{Gladstone2009}  find that the highest quality ULX  spectra can be modeled by a cool disc component together with a power law which breaks/rolls over above 3 keV.   The spectrum of \hxi\ is significantly harder than any of the sources studied by \cite{Gladstone2009}.   It is similar to the spectra of a sample of extreme ULX  (L$_{X} > 5\times10^{40}$ ergs s$^{-1}$ studied by \cite{Sutton2012}.     \cite{Sutton2012} suggest that extreme ULX with relatively hard spectra may be black hole binaries in the low, hard state (see also \cite{Feng2006}).  If we make the assumption that \hxi\ is a black hole in the low state radiating at less than  10\% of the Eddington luminosity, then the mass of the black hole is greater than 7600$M_{\odot}$, putting it in the ``Intermediate Mass Black Hole" category.   

 Alternately, \hxi\ could be a highly super-Eddington massive stellar BH (M$\sim$100 M$_{\odot}$ radiating at $\sim$10 times Eddington).    Recently \cite{Sutton2013} suggested that  most ULXs can be explained as stellar-mass black holes accreting at and above the Eddington limit.  Such high accretion rates leads to the formation of a wind or jet.  The X-ray spectrum and variability characteristics depend on the accretion rate and inclination of the system.  The ``hard'' ultra luminous state occurs when observing primarily down the jet funnel to the hard central engine.  A softer component is observed from the wind at higher inclination angles.    It is impossible to distinguish between these two possibilities (IMBH radiating at ~10\% Eddington vs. super Eddington accretion onto a 100$M_{\odot}$ black hole) with the current data: high quality X-ray spectra (several thousand counts) are required to do the spectral fitting.  This would require extremely long exposures for \chandra\ since other observatories cannot resolve \hxi\ and \hxii..  
  
  If  \hxi\ is an IMBH in a low accretion state, it has the potential to transition to a high state.  If \hxi\ transitions to a high state, it will  reach luminosities which will classify it as a Hyper Luminous X-ray source (\hlx\ defined as  L$_{X}\ge10^{41}$ ergs s$^{-1}$ \cite{Gao2003}).      It is potentially significant that \haroii\ is a low metallicity galaxy, apparently similar to higher redshift objects.  An \imbh\ may be a ``seed''  black hole which is currently growing rapidly to form a supermassive black hole \citep{Jia2011}.    \cite{McHattie2014}  note that the HI line width is consistent with the rotation curve measured by \cite{Ostlin1999}.  This suggests that the HI resides at the dynamical center of the galaxy, where one would expect to a find or grow a supermassive black hole. 
  
   The ULX in I~Zw~18 is another excellent candidate for an IMBH in a low metallicity starburst \citep{Kaaret2013}.    It is also interesting to note that the HLX ESO 243-49 HLX-1 is located in the outskirts  of an S0 galaxy, and may be a the stripped nucleus of an accreted dwarf galaxy \citep[][and references therein]{Farrell2011}.     \haroii\ is precisely  the type of system which could be the progenitor of a ``stripped dwarf nucleus''.  Finally, we note that  there is good evidence that \ulx\ are more common in low metallicity dwarf galaxies (e.g. \cite{Mapelli2010,Prestwich2013, Brorby2014}).   In particular,  \cite{BasuZych2013}  find that the 2-10 keV X-ray luminosity per unit star formation rate in a sample of low redshift LBAs is elevated relative to near-solar metallicity galaxies.  They attribute the excess as excess ULX in these metal poor systems.  It is likely that \haroii\ is not unique, and that extreme ULX/HLX are common in LBAs.   

 \section{The Significance of \hxii: XRB and Feedback in LAE and LBAs}

The X-ray luminosity of \hxii\  is characteristic of a \ulx.  The X-ray spectrum (soft power law $\Gamma=$2.2) is  also consistent with the interpretation of \hxii\ as a \ulx\ \citep{Gladstone2009}.   We note that most \ulx\ with soft spectra have X-ray luminosities $<2\times 10^{40}$ ergs s$^{-1}$, making \hxii\ one of the most luminous soft \ulx\ known.   Based on the X-ray characteristics, and it's association with young (age $\sim$ 10 Myr, \cite{Adamo2010}) star clusters in Knot C,  we conclude that \hxii\ is a stellar mass black hole (M$\sim$10-100$M_{\odot}$, \cite{Zampieri2009}) in a young High Mass X-ray Binary.  

Recent studies of \ulx\ with high quality spectra have shown that a subset of these sources have a soft X-ray excess and a turnover in the spectrum above 3 keV \citep{Gladstone2009}.    These sources can be explained in terms of super-Eddington accretion.   In this scenario,  a soft ultraluminous spectrum  is  viewed (at least in part) through a massive outflowing wind, with the softness of the spectrum due to the Compton down scattering of inner-disc photons in this very optically thick wind \citep{Sutton2013}.     The wind is  a direct consequence of super-Eddington accretion - the intense radiation release of the inner-disc results in a radiatively-driven wind from the loosely bound surface of the accretion disc, that itself becomes geometrically thick in its inner regions \citep{Poutanen2007,Middleton2015}.    

We suggest that \hxii\ is a \ulx\ accreting in the super-Eddington state.  The X-ray spectrum (soft power law $\Gamma=$2.2) is consistent with the interpretation of \hxii\ as a super Eddington accretor  \citep{Gladstone2009}.   However, we note that most \ulx\ with soft spectra have X-ray luminosities $<2\times 10^{40}$ ergs s$^{-1}$.  The  X-ray luminosity of \hxii\  ($5\times 10^{40}$ ergs s$^{-1}$) is too high for a ``standard'' stellar mass black hole (M$<$20 $M_{\odot}$).   It seems likely 
requires that the black hole have a mass in the range 20-100$M_{\odot}$.

The importance of outflows/feedback in allowing \lya\ to escape  is widely accepted \citep{Heckman2011,Wofford2013, Orsi2012}.  However,  the earliest  source of the outflow is assumed to be winds from supernova and high mass stars \citep[see evolutionary scenario in][]{Mas-Hesse2003}  At later times, an AGN can form and  also contribute.    The association of \hxii\ with the \lya\ emission raises the possibility that winds from the compact object may play a significant role in sweeping up the neutral gas and allowing the \lya\ to escape.  The mechanical luminosity in the wind of an XRB is generally comparable to the radiative luminosity \citep{Gallo2005,Justham2012}:  there is evidence for some systems that the mechanical luminosity might dominate the radiative luminosity by several orders of magnitude \citep{Pakull2010}.     

 \begin{figure}
\includegraphics{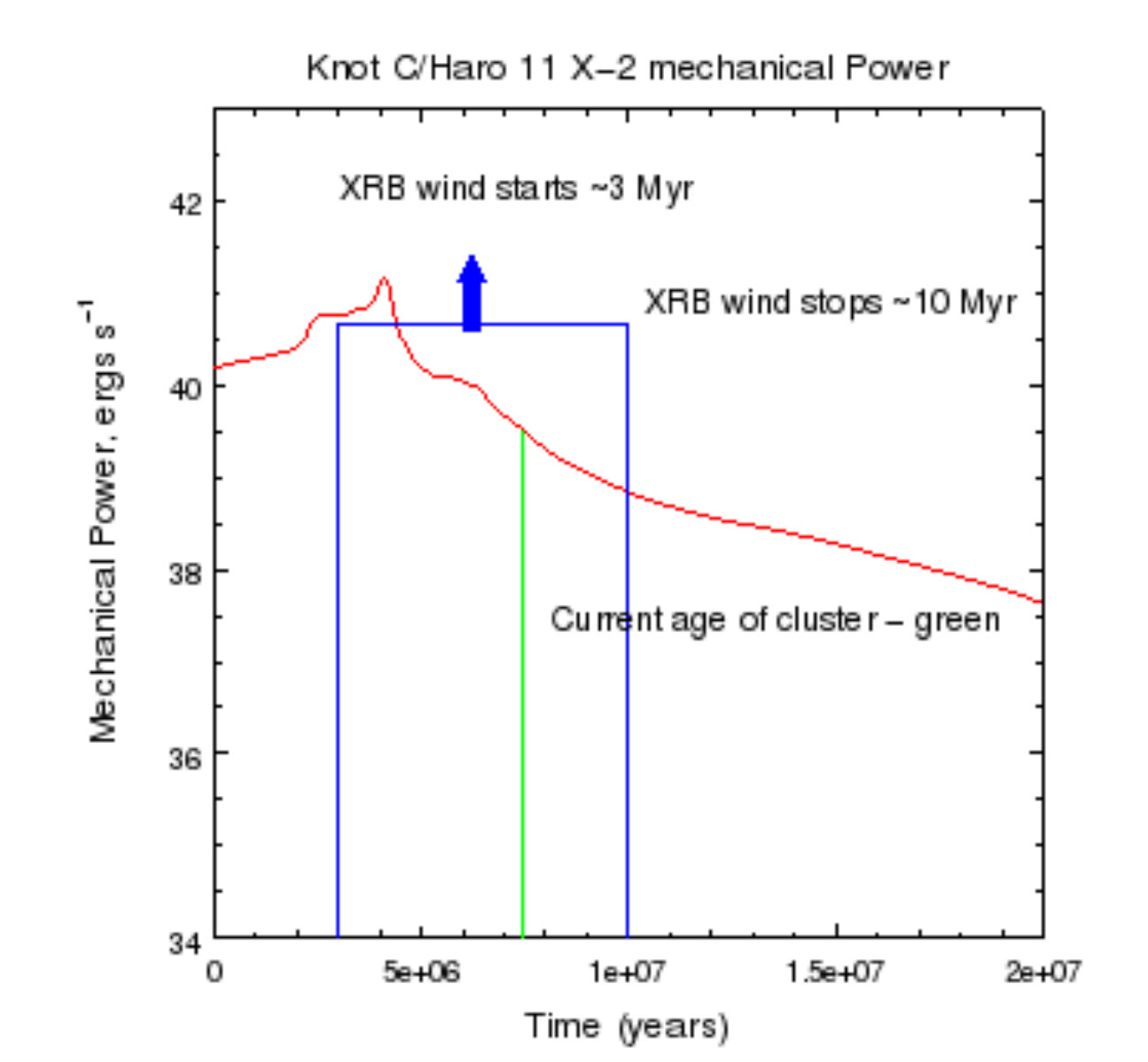}
\caption{Mechanical power vs. time for Knot C and \hxi.  .  The red curve shows the mechanical power (supernovae plus stellar winds) from a 10$^{7}$ M$_{\odot}$ instantaneous burst of star formation.   The age of Knot C ($\sim$ 10 Myr, \cite{Adamo2010}) is shown in green.  The blue shows the approximate mechanical power from a 5$\times$10$^{40}$  ergs s$^{-1}$ ULX, assuming (1) that the mechanical power $\sim$ radiative luminosity (2) the outflow turns on when the ULX is 3 Myr old and turns off at 10 Myr.   The mechanical power from a ULX may be greater than the radiative luminosity, so the blue line is a lower limit as indicated by the  arrow.}
\label{fig:pow}

\end{figure}

Figure~\ref{fig:pow} shows the mechanical power from supernovae and stellar winds  as a function of time for the starburst region Knot C (red curve) compared to estimated feedback from a ULX (blue curve).  We use the  Starburst99 code \citep{Leitherer1999, Vazquez2005, Leitherer2010} to model the mechanical power (supernova plus stellar winds) for an instantaneous starburst of fixed mass.   We adopt a \cite{Kroupa2001} Initial Mass Function (IMF)  with mass limits 0.1 and 100 M$_{\odot}$, a turnover stellar mass at 0.5 MM$_{\odot}$ and lower/upper IMF exponents of $\alpha$ =1.3/2.3.   The input star  cluster mass is 10$^{7}$ M$_{\odot}$  and metallicity $Z=0.004$  (values are those derived for Knot C by \cite{Adamo2010}).    The blue line shows  the mechanical power from the ULX, with the conservative assumption that the mechanical power is approximately equal to the radiative power.  The wind from the ULX could turn on $\sim$3 Myr after formation, and last for the lifetime of the binary ($\sim$10 Myr: \cite{Rappaport2010}).    The age of the cluster (estimated by \cite{Adamo2010}) is shown in green.   It is clear that a single ULX has the potential to contribute significantly  to, and possibly dominate,  the mechanical power.


\section{Summary and Conclusions}

We find two point sources embedded in the diffuse emission in the \lya\ galaxy \haroii.

\hxi\  is extremely luminous, spatially compact, has a hard X-ray spectrum and is associated with Knot B and a compact HI absorption source.     There is no evidence for a LLAGN in the optical or infrared emission lines, and no high surface brightness radio core.  We cannot rule out an AGN whose optical and IR signatures are diluted by the intense starburst.   We conclude that this source is most likely  dominated by a single luminous XRB on the basis of energetics and the hard X-ray spectrum, similar to the IMBH candidate in M82 (M82 X-1).   We cannot distinguish between a stellar black hole (M$\sim$100 M$_{\odot}$) and an  IMBH (M$\sim$7600 M$_{\odot}$)  with current data.  If \hxi\ is indeed dominated by a single \bhb, the hard spectrum suggests it is an \imbh\ in a low accretion state.   An \imbh\ may be a ``seed''  black hole which is currently growing rapidly to form a supermassive black hole \citep{Jia2011}. 

\hxii\ is a \ulx, and is most likely a young \hmxb.  It is associated with Knot C, and is coincident with the  \lya\ line emission.  We suggest that winds from the \ulx\ may play a significant role in sweeping up the neutral gas and allowing the \lya\ to escape.   \haroii\ is an \lyba, and hence similar to \lbg\ at high redshifts.  This raises the possibility that  feedback from \ulx\ play  an important role in  allowing \lya\ emission to escape from low metallicity galaxies in the early universe.





\section{Acknowledgements}

The scientific results reported in this article are based on  data obtained from the Chandra Data Archive.   This research has made use of software provided by the Chandra X-ray Center (CXC) in the application packages CIAO, ChIPS, and Sherpa.   

Based on observations made with the NASA/ESA Hubble Space Telescope, and obtained from the Hubble Legacy Archive, which is a collaboration between the Space Telescope Science Institute (STScI/NASA), the Space Telescope European Coordinating Facility (ST-ECF/ESA) and the Canadian Astronomy Data Centre (CADC/NRC/CSA).

\bibliographystyle{apj}
\bibliography{Haro11}




\end{document}